\documentclass[journal]{IEEEtran}
\usepackage{graphicx,amssymb,amsmath}
\usepackage{multicol}
\usepackage[noadjust]{cite}
\usepackage{setspace}
\usepackage{stfloats,subfigure}
\usepackage{midfloat}
\usepackage[normal]{threeparttable}
\usepackage{amsthm}
\usepackage{cuted}
\usepackage{cases,subeqnarray}
\usepackage{bm,multirow,bigstrut}
\usepackage{textcomp}
\usepackage{latexsym,bm}
\usepackage{booktabs,changebar}
\usepackage{xcolor}
\usepackage{mathtools}
\usepackage{dsfont}
\usepackage{extarrows}
\usepackage{epstopdf}   
\theoremstyle{plain}

\theoremstyle{plain}

\IEEEoverridecommandlockouts
\begin{document}

\title{Novel Device-to-Device Discovery Scheme based on Random Backoff in LTE-Advanced Networks}

\author{Jiayi~Zhang,~\IEEEmembership{Member,~IEEE}, Likai Deng, Xu Li, Yuchuan Zhou, Yanan Liang, and Ying Liu
\thanks{Copyright (c) 2015 IEEE. Personal use of this material is permitted. However, permission to use this material for any other purposes must be obtained from the IEEE by sending a request to pubs-permissions@ieee.org.}
\thanks{This work was supported in part by the National Natural Science Foundation of China (Grant No. 61601020) and the Fundamental Research Funds for the Central Universities (Grant Nos. 2016RC013, 2017JBM319 and 2016JBZ003).}%
\thanks{The authors are with the School of Electronics and Information Engineering, Beijing Jiaotong University, Beijing 100044, P. R. China (e-mail: jiayizhang@bjtu.edu.cn).}
}

\maketitle

\begin{abstract}
Device-to-Device (D2D) discovery is a key enabler of D2D communications for the direct exchange of local area traffic between proximity users (UEs) to improve spectral efficiency. The direct D2D discovery relies on the capabilities of the D2D UEs to autonomously indicate their presence to proximity D2D UEs. {Despite} its potential of reducing energy and signalling burden, the direct D2D discovery {has not drawn adequate attention}. In this paper, we propose a direct D2D discovery scheme based on the random backoff procedure, where D2D UEs randomly choose a backoff interval and retransmit a beacon. Compared with existing schemes, the performance of the proposed scheme can be significantly enhanced in terms of the discovery probability and the discovery delay. Several useful guidelines for its design are proposed based on our analysis. Finally, numerical results provide valuable insights on the performance tradeoff inherent to the proposed D2D discovery scheme.
\end{abstract}


\IEEEpeerreviewmaketitle

\section{Introduction}
With the accelerating growth of context-aware applications, the device-to-device (D2D) communication plays an important underlay role over the cellular network \cite{wong2017key}. However, D2D communication cannot be realized without an efficient D2D discovery scheme, which involves nearby users (UEs) to search for the proximity-based service. {Despite} its importance, relatively less attention and research have been addressed to D2D discovery.


Generally speaking, {a pair of UEs moving within close proximity to each other in {an LTE-A network}, can establish a D2D link with or without the assistance of the serving eNB(s) \cite{liu2014device}.} More specifically, the D2D discovery scheme can be divided into two types: network-assisted and direct discovery \cite{xenakis2016performance}. The network-assisted D2D discovery requires the D2D UE connected to the BS all the time, which leads to inefficient use of network resource, large battery consumption, and the challenging of stability. On the contrary, the direct D2D {discovery} relies on the abilities of the D2D UEs to autonomously discover other D2D UEs. The BS only periodically broadcasts the available resources for transmitting and receiving discovery beacons, by which the energy and signaling overhead at the BS can be hugely reduced. Moreover, the direct D2D {discovery} has been employed in experimental prototypes \cite{wu2013flashlinq}. In this paper, we focus on the performance of the direct D2D {discovery} in LTE-A networks.

For the network-assisted D2D discovery, a social-aware D2D discovery scheme underlying cellular networks was proposed in \cite{zhang2015social}, where each group of D2D UEs is allocated the optimal beacons probing rate with constant intervals. However, social network characteristics (such as real human mobility traces) are {difficult} to be acquired. Moreover, the D2D UEs are typically power-limited, which means the energy-consuming D2D discovery scheme is not suitable for such problem. Some works focus on adapting the probing frequency to reduce energy consumption and improve network performance \cite{drula2007adaptive,vigato2011joint}. {Another key point of the direct D2D discovery is to reduce collision. Currently, greedy-selection based discovery methods are widely proposed \cite{simsek2013device,jung2014discovery}. Greedy selection requires UEs to scan the whole resource before selecting the most optimistical resource, which can avoid some interference in a degree. However, collisions cannot be averted when UEs are very close to each other. Since the load on D2D network varies significantly in both time and location, many researchers applied stochastic geometry model to capture network dynamics when evaluating D2D discovery performance.} {Different from these works investigating the link-level D2D discovery where the metric is signal-to-noise ratio (SNR), we evaluate the system-level performance for the direct D2D discovery}.

In contrast to the network-assisted D2D discovery, little research has been addressed on the topic of the direct D2D {discovery} scheme. The direct D2D discovery has the similar purpose of the random access procedure, {which establishes a connection to the BS and avoid a collision of UEs.} {In LTE-A network, each fixed-length radio frame consists of multiple sub-frames. Random access procedures are restricted to specific sub-frames referred as random-access slots.} The random backoff method has been widely utilized in the random access procedure to combat network collisions \cite{he2013semi,misra2015semi}. Herein, {we propose} a direct D2D discovery scheme based on the random backoff strategy to avoid collisions. {The proposed scheme renders an increase in} terms of D2D discovery ratio compared with the existing schemes. More specifically, all D2D UEs randomly selects a resource block (RB) to send a preamble to nearby UEs in a distributed discovery manner. If two or more D2D UEs select the same RB, the D2D discovery fail because of a collision. In the following, these failed UEs will choose a random backoff window size and retransmit in a RB to build connections. The proposed D2D discovery scheme fully utilize the merit of the random backoff procedure, which, to the best of our knowledge, has not been taken into account in previous works.

The main contributions can be summarized as follows:
\begin{itemize}
\item We establish a new paradigm for D2D discovery with the help of the random backoff method. Our proposed scheme can readily be utilized to current LTE-A network without significant modification.
\item Based on this paradigm, we propose an efficient random backoff aware D2D discovery scheme, in which D2D UEs of collisions resend the beacons in next random allocated RB. With new probabilistic ``tools", Balls \& Bins combinatorics model \cite{wei2015modeling}, we analyze the performance in terms of the average number of discovered D2D UEs, the collision probability, and the D2D discovery delay.
\item The simulation results validate that our scheme can significantly increase the average number of discovered D2D UEs, and can also reduce both the D2D discovery delay and the collision probability compared to the existing scheme \cite{R1140841}.
\end{itemize}


\section{System Model}\label{se:system_model}

We consider {an LTE-A} cellular network, where the radio resource is divided into RB in time and frequency domain. For concreteness, the RBs specified in 3GPP standard \cite{3gppTS36211} is considered in the rest of the paper. Each RB includes 12 OFDM symbols and one slot. A D2D Discovery Zone (D2D DZ) is designed for D2D discovery, and appears periodically every $T_d$ s. There are $2R$ RBs in a DZ for D2D UEs to send discovery beacons in $T_{DZ}$ s. Thus, the discovery period is denoted by $T_d+T_{DZ}$.

{Similar to \cite{R1140841}, we consider a general case wherein each discovery resource comprises of 2 RB-pairs, thus there are $R$ discovery resources in each DZ. D2D UEs randomly select a discovery resource to transmit discovery beacons in each coming DZ. If only one UE occupies a discovery resource (i.e. a pair of RB),} the D2D discovery is successful under the assumption that one RB-pair is needed for sending a beacon. Otherwise, the D2D discovery procedure is failed if two or more UEs select the same RB-pair. The unsuccessful UEs then retransmit their discovery beacons in the next DZ for maximum $L_{\max}$ times. In each DZ, the number of D2D UEs ${M_i}$ can be expressed as
\begin{align}
{M_i} = \sum\limits_{l = 1}^{{L_{\max }}} {{M_i}[l]},
\end{align}
where ${{M_i}[l]}$ denotes the number of D2D UEs sending beacons for the $l$th time in the $i$th DZ ($DZ(i)$). {Note that ${M_i}[1]$ is the number of new UEs which request to establish D2D communication links.} Without loss of generality, we assume the number of D2D UEs which carry out the beaconing procedure to discover potential peers follows the beta distribution
\begin{align}\label{beta_distribution}
p(t) = \frac{{{t^{\alpha  - 1}}{{(T - t)}^{\beta  - 1}}}}{{{T^{\alpha  + \beta  - 1}}B(\alpha ,\beta )}},\;\;\;\alpha>0,\;\beta>0,
\end{align}
where $B(\alpha,\beta)$ is the Beta function and $T$ is a time interval. {The beta distribution has been suggested to model the M2M overload traffic distribution by 3GPP \cite{wu2013fasa}.} Then, the number of new arrived D2D UEs in $DZ(i)$ is given by
\begin{align}\label{new_UE}
{{M_i}\left[1\right]} = \left\{ \begin{array}{l}
 M\int_{0}^{T_d} {p\left(t\right)dt}, \begin{array}{*{20}{c}}
{}&{i = 1\qquad }
\end{array}\\
\\
M\int_{\left(i - 2\right){T_{DZ}}+\left(i - 1\right){T_d}}^{\left(i - 1\right){T_{DZ}}+i{T_d}} {p\left(t\right)dt}., \begin{array}{*{20}{c}}
{}&{i \geq 2}
\end{array}
\end{array} \right.
\end{align}
{where $M$ is the total number of D2D UEs in network. The upper and lower limits in (3) is under the defination that new UEs in $DZ(i)$ arrive between the start time of $DZ(i-1)$ and the start time of $DZ(i)$. More specifically, we assume the time period is $T_d$ when $i=1$.}

In order to derive the collision probability, let us focus on a given DZ which includes $R$ RB-pairs. A collision emerges when two or more UEs select the same RB-pair. Assuming the number of D2D UEs is far larger than the $R$, it is easy to observe that the number of successful D2D discovery UEs is in most generality a random variable $S$ taking values in the range 0 to $R$. This kind of problem can be cast into a Balls \& Bins combinatorics model \cite{wei2015modeling}. Moreover, let ${M_{i,S}}[l]$ be the successful D2D discovery UEs after the $l$th transmission in $DZ(i)$, and ${M_{i,F}}[l]$ be the failed D2D discovery UEs after the $l$th transmission in $DZ(i)$. According to \cite[Theorem II.3]{wei2015modeling}, ${M_{i,S}}[l]$ and ${M_{i,F}}[l]$ {can respectively} given by
\begin{align}
{M_{i,S}}[l] &= {M_i}[l]*{(1 - \frac{1}{R})^{{M_i}[l] - 1}},\label{new_UE_S} \\
{M_{i,F}}[l] &= {M_i}[l] - {M_{i,S}}[l].\label{new_UE_F}
\end{align}

\vspace{-0.4cm}
\section{D2D Discovery Based on Random Backoff}\label{se:scheme}
\subsection{Random Backoff Scheme}
\begin{figure}[htbp]
\centering
\includegraphics[scale=0.55]{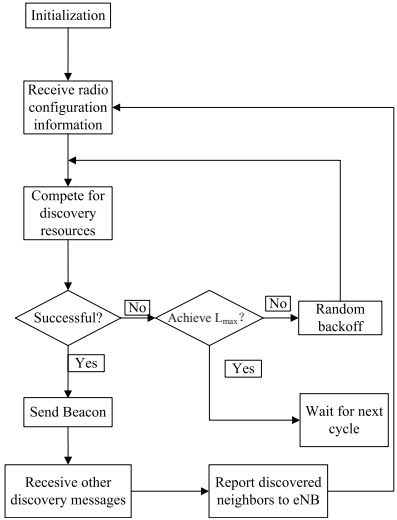}\\
\caption{Flow chart of random-backoff based D2D discovery scheme.}
\label{Flowchart}
\end{figure}

{As shown in Fig. \ref{Flowchart}, when D2D UEs register in a cell and get radio configuration information from eNB's broadcast, they will start D2D discovery procedure. In each DZ, the D2D UEs including both new arrived UEs and the failed UEs in previous DZs send their beacons for discovery. In contrast to existing schemes, the failed UEs will randomly select a backoff period before their next beacons transmission. Moreover, each failed UE will check whether the retransmission
time achieves the predefined maximum value before random backoff. If the retransmission time achieves its maximum value, the D2D UE has to quit competing and wait for the next discovery cycle. Thanks to the proposed scheme, the collision probability can be effectively decreased due to the reduced number of D2D UEs in one DZ.}
\subsection{Number of Discovered UEs}
{More specifically, when a D2D UE fails to send beacon in $DZ(g)$ for the ($l-1$)th time, it will randomly choose a window length of $W_{BO}$ $(1 \leq W_{BO} \leq W)$ to backoff, and then retransmit its beacon in the next coming DZ. If $\eta$ percentage of these D2D UEs selects $DZ(i)$ for the $l$th transmission, which means $\eta$ percentage of these D2D UEs chooses the backoff window length of $(i-g-1)$.} The number of UEs for the $l$th retransmission in each DZ is given by
\begin{align}\label{UE_l}
{M_i}[l] = \sum\limits_g {\eta {M_{g,F}}[l - 1]},
\end{align}
where ${M_{g,F}}[l - 1]$ denotes the number of failed UEs for the $(l-1)$th time transmission at $DZ(g)$. Substituting (6) into (4), we can easily derive $M_{i,S}[l]$ ($l \geq 1$). Without loss of generality, we assume the random backoff length follows the uniform distribution as $\eta=1/W$. {Note that the number of UEs in each DZ includes the failed UEs in previous DZs and new arrived UEs.} Then, the total number of UEs in $DZ(i)${, defined as $M_i$}, is written as
\begin{align}\label{UE_i}
 {M_i = M_i[1] + \eta\cdot\sum\limits_{w=1}^{W}\sum\limits_{l=2}^{L_{max}}(M_{(i-w),F}[l-1])}.
\end{align}
 {where $w$ denotes the random backoff window. From \eqref{UE_i} we can see that the formula is only applicable to the case when $i>W$. For simplicity, we assume that retransmission will not happen when $i\leq W$.} Then with the theory of discrete-time Markov chain, the total number of successful UEs after $K$ DZ can be expressed as
\begin{align}\label{UE_K}
{{M_K}(S)} = \left\{ \begin{array}{l}
\sum\limits_{i = 1}^W {{M_{i,S}}[1]}, \begin{array}{*{20}{c}}
{}&{i \le W\qquad }
\end{array}\\
\\
\sum\limits_{i = 1}^K {\sum\limits_{l = 1}^{{L_{\max }}} {{M_{i,S}}[l]}. \begin{array}{*{20}{c}}
{}&{W \le i \le K}
\end{array}}
\end{array} \right.
\end{align}

\subsection{Delay Analysis}
The average discovery delay ${{\overline D} _S}$ is defined as the ratio between the total access delay for all of the successfully discovered UEs and the total number of the successfully discovered UEs, and is given by \cite[Eq. (33)]{wei2015modeling}
\begin{align}\label{UE_delay}
{{\overline D}_S} = \frac{{\sum\limits_{i = 1}^K {\sum\limits_{l = 1}^{{L_{\max }}} {\left({M_{i,S}}\left[l\right]{\Delta _{l,i}}\right)} } }}{{\sum\limits_{i = 1}^K {\sum\limits_{l = 1}^{{L_{\max }}} {{M_{i,S}}\left[l\right]} } }},
\end{align}
where ${\Delta _{l,i}}$ represents the average access delay of discovered UEs after the $l$th transmission in $DZ(i)$ as
\begin{align}\label{UE_time}
{\Delta _{l,i}} = {\overline T}_{ar}+ \left(l - 1\right){\overline T}_{bo} + {T_h},
\end{align}
where $T_h$ is the response and processing time of UEs, ${\overline T}_{bo}$ denotes the average waiting time required by an UE to perform backoff and retransmit a beacon as
\begin{align}\label{T_bo}
{\overline T}_{bo}  = \frac{1}{W}\sum\limits_{{W_{BO}} = 1}^W {{W_{BO}}\left( {T_d + {T_{DZ}}} \right)}
\end{align}
and ${\overline T}_{ar}$ is the delay of new arrived UEs in $DZ(i)$, which is given by
\begin{align}\label{T_ar}
{{\overline T}_{ar} \left[i\right]} = \left\{ \begin{array}{l}
 \int_{0}^{T_d} {t\cdot p\left(t\right)dt}, \begin{array}{*{20}{c}}
{}&{i = 1\qquad }
\end{array}\\
\\
\int_{\left(i - 2\right){T_{DZ}}+\left(i - 1\right){T_d}}^{\left(i - 1\right){T_{DZ}}+i{T_d}} {t\cdot p\left(t\right)dt}. \begin{array}{*{20}{c}}
{}&{i \geq 2}
\end{array}
\end{array} \right.
\end{align}
{The upper and lower limits in (12) is under the definition that new UEs in $DZ(i)$ arrive between the start time of $DZ(i-1)$ and the start time of $DZ(i)$. More specifically, we assume the time period is $T_d$ when $i=1$. Moreover, the average delay of all new arrived UEs in the DZ can be derived from \eqref{T_ar}} as
\begin{align}\label{UE_avearge_delay}
{\overline T}_{ar}  = \frac{1}{K}\sum\limits_{i = 1}^K {{\overline T}_{ar} \left[ i \right]}.
\end{align}
Note that ${T_d} \gg {T_{DZ}}$, {\eqref{T_ar} }can be further reduced to
\begin{align}\label{UE_avearge_delay_reduce}
{\overline T}_{ar} \left[i\right]  &= \int_{\left( {i - 1} \right){T_d}}^{i{T_d}} {t \cdot p\left( t \right)d t} \notag \\
&= - 9.21 \times {10^{ - 6}} \times {i^6} + 5.28 \times {10^4} \times {i^5} - 0.0125{i^4} \notag \\
&+ 0.0985{i^3} - 0.13514{i^2} + 0.086i - 0.021.
\end{align}
Substituting {\eqref{UE_time}, \eqref{T_bo} and \eqref{UE_avearge_delay}} into \eqref{UE_delay}, we can derive the average discovery delay for the successfully discovered UEs.

\section{Numerical Results}\label{se:numerical_results}
In this section, we present numerical results for the proposed random backoff-assisted direct D2D discovery scheme. Several useful guidelines for the design of the proposed D2D discovery are derived. In both simulation and analysis, the bandwidth for the LTE-A system equal to 10 MHz, and the Beta distribution is utilized to model D2D UEs. For beta-distributed arrivals, we choose the shape parameters as $\alpha = 3$ and $\beta = 4$. Similar to LTE-A specifications in \cite{3GPPTS37868}, key simulation assumptions for D2D discovery can be found in Table \ref{table1}.
\begin{table}[tb]
\renewcommand{\thetable}{\Roman{table}}
\caption{Key Parameters for Simulations}
\label{table1}
\centering
\begin{tabular}{|c|c|c|}
\hline
Notation  & Definition & Values\\
\hline
$M$& Total number of D2D UEs & 450   \\
\hline
$R$ & Number of PRB-pairs & 22   \\
\hline
$T_{DZ}$ & Length of DZ & 30 ms   \\
\hline
$T_{d}$ & Interval of two DZs & 10 s   \\
\hline
$K$ & Number of DZs & 20   \\
\hline
$L_{\max}$ & Maximum number of retransmission & variable   \\
\hline
$T_h$ & Response and processing delay & 5 ms  \\
\hline
$W$ & Random backoff window length & 1-6   \\
\hline
\end{tabular}
\end{table}
\subsection{Average Number of Discovered UEs}
We first investigate the average number of discovered UEs for the proposed random backoff-assisted D2D discovery scheme. If the PRB-pair is occupied by only one UE, the D2D discovery procedure is successful. According to \cite{3GPPTS37868}, the failed probability of UEs resending their beacons for the $l$th time is defined as $P_F = 1/{e^l}$, and the successful probability of UEs resending their beacons for the $l$th time is given by ${P_S} = 1 - \frac{1}{{{e^l}}}$. With the help of \eqref{UE_i}, the discovered UEs in $DZ(i)$ can be expressed as
\begin{align}\label{UE_discovered}
 {{M_{i,S}} = \sum\limits_{l = 1}^{L_{max}}{ {M_{i,S}}\left[l\right]\cdot P_S}.}
\end{align}

\begin{figure}[tbp]
\centering
\includegraphics[width=0.8\linewidth]{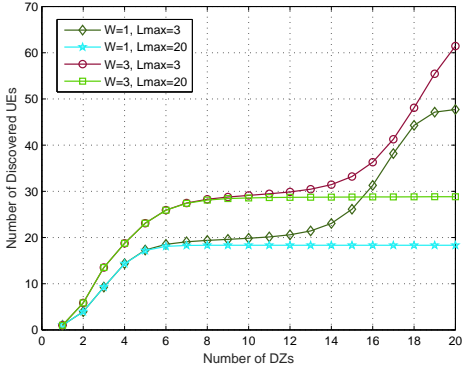}
\caption{Number of discovered UEs for the proposed random backoff-assisted D2D discovery scheme.}
\label{diff_W_Lmax}
\end{figure}

The impact of different D2D discovery scheme on the number of discovered UEs has been investigated in Fig. \ref{diff_W_Lmax}. Four kinds of the D2D discover scheme are considered: 1) the case of $W=1$ and $L_{\max}=3$ denotes the maximum number of retransmission is 3 and without the random backoff scheme; 2) the case of $W=1$ and $L_{\max}=20$ denotes the failed UEs can choose every DZ to send their beacons without random backoff scheme; 3) the case of $W=3$ and $L_{\max}=3$ denotes the maximum number of retransmission is 3 and the random backoff window is 3; 4) the case of $W=3$ and $L_{\max}=20$ denotes the maximum number of retransmission is unlimited and the random backoff window is 3.

It is clear from Fig. \ref{diff_W_Lmax} that our proposed random backoff-assisted D2D discovery scheme (case 3) can obtain the maximum number of discovered UEs as 61.4, and the D2D discovery probability as 13.6\%. Compared with previous results \cite{R1140841}, where the D2D discovery probability is only 6.4\%, the proposed scheme can effectively improve discovery probability. This follows from the fact that the collision probability is reduced by allowing less D2D UEs to send beacons in a DZ. Moreover, the number of discovered UEs for the case with the random backoff scheme increases 20-50\% compared with the one without the random backoff scheme. It is also interesting to find that the case under the limit of maximum retransmissions can achieve larger number of discovered UEs, since less collisions tend to happen in a DZ for our scheme.

\begin{figure}[tbp]
\centering
\includegraphics[width=0.8\linewidth]{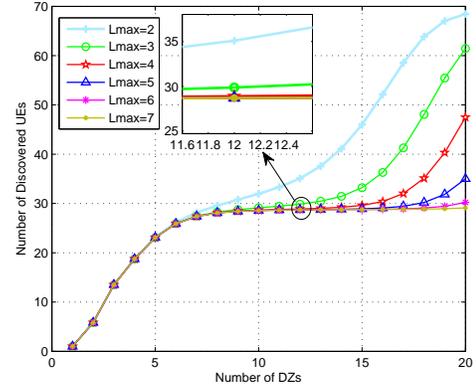}
\caption{Number of discovered UEs against different $L_{\max}$ for the proposed random backoff-assisted D2D discovery scheme.}
\label{diff_Lmax}
\end{figure}

\begin{figure}[tbp]
\centering
\includegraphics[width=0.8\linewidth]{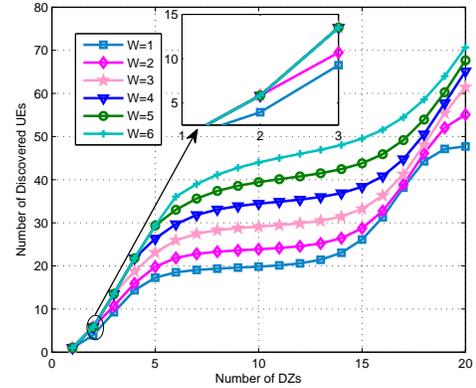}\\
\caption{Number of discovered UEs against different $W$ for the proposed random backoff-assisted D2D discovery scheme.}
\label{Us_W}
\end{figure}

In Fig. \ref{diff_Lmax}, we further {show} the D2D discovery number as a function of the maximum number of retransmission $L_{\max}$ with $W=3$. As expected, the number of discovered UEs decreases with $L_{\max}$. Moreover, the gap between the corresponding curves decreases as $L_{\max}$ increases which implies that its effect becomes less pronounced. {Moreover, we can see that the lines overlap together in first DZs, which is related to the value of the maximum number of retransmission $L_{\max}$. In first DZs, the discover probability is high due to the small number of new arrived D2D UEs. However, with the number of arrived D2D UEs, the collision probability increases. When $L_{\max}$ increases from 2 to 7, the congested UEs increase gradually. When the number of congested UEs accumulates to a certain value, the number of discovered UEs does not increase any more.}

The effect of the random backoff window $W$ on the system performance is revealed in Fig. \ref{Us_W}. It is clear to see that the collision probability can be decreased by increasing $W$. This is due to the fact that the number of UEs in one DZ is reduced. {Since we assume that the retransmission is not allowed in first $W$ DZs, the number of discovered UEs in first $W$ DZs is constant.} The benefit of large values of $W$ is more pronounced when the number of DZ increases from 8 to 15. Therefore, we can  achieve a target D2D discovery probability by adjusting the values of $W$ and $L_{\max}$.

\subsection{Average Delay}

\begin{figure}[tbp]
\centering
\includegraphics[width=0.8\linewidth]{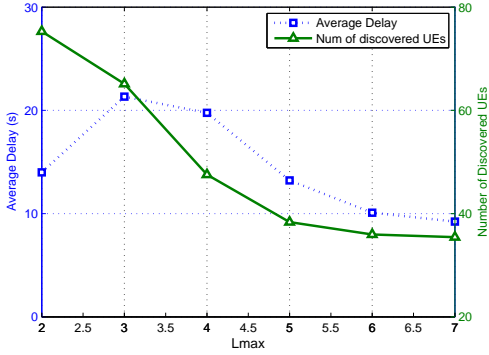}\\
\caption{Average delay ${\overline D}_S$ against different $L_{\max}$ for the proposed random backoff-assisted D2D discovered scheme.}
\label{delay_L}
\end{figure}
\begin{figure}[tbp]
\centering
\includegraphics[width=0.8\linewidth]{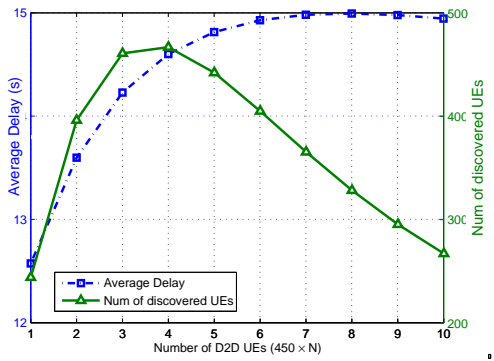}\\
\caption{Average delay ${\overline D}_S$ against different number of D2D UEs for the proposed random backoff-assisted D2D discovered scheme.}
\label{delay_massive_L}
\end{figure}
Fig. \ref{delay_L} investigates the average delay ${\overline D}_S$ of the proposed D2D random backoff-assisted discovered scheme as a function of $L_{\max}$. It is readily seen that the average delay ${\overline D}_S$ firstly increases from $L_{\max}=2$ to $L_{\max}=3$, while ${\overline D}_S$ reduces as $L_{\max}$ becomes larger. As anticipated, the average number of discovered UEs is a monotonically decreasing function in $L_{\max}$, which is in line with Fig. \ref{diff_Lmax}. From Fig. \ref{delay_L}, we can find that the decreasing of the average delay ${\overline D}_S$ is obtained by reducing the number of discovered UEs. Furthermore, in order to more closely reveal the tradeoff between the average delay ${\overline D}_S$ and the average number of discovered UEs ${M_i}\left( S \right)$, we introduce a reasonable metric as $\eta_S = {\overline D}_S/{M_i}\left( S \right)$ which can be used to decide the optimum number of retransmission for the proposed scheme. It is worth mentioning that the minimum value of $\eta_S=0.186$ is achieved by setting $L_{\max}=2$.

Furthermore, we consider a large period of $K=200$ DZs and massive D2D UEs. In Fig. \ref{delay_massive_L}, the effect of the number of UEs on the average delay and the number of discovered UEs is illustrated. It is clear to see that the maximum number of discovered UEs is 467 when $M=1800$ ($N=4$). However, the average delay increases as $M$ increases which implies that the collision probability becomes large. When $M \ge 2250$ ($N\ge 5$), the average delay increases to a constant value, while the number of discovered UEs still decreases. This is anticipated from the definition of the average delay in \eqref{UE_delay}.

\section{Conclusion}\label{se:conclusion}
In this paper, we propose a novel D2D discovery scheme based on the random backoff procedure. The proposed scheme can readily be utilized in the current LTE-A system with trivial modification, since it is designed based on the existing D2D discovery procedure and random backoff method. Furthermore, we derive analytical expressions for the average discovered UEs and the average delay, respectively. Based on our analysis, the optimum number of retransmissions and number of random backoff window can be obtained to achieve a target D2D discovery probability. Extensive simulations show that our proposed scheme can significantly increase the D2D discovery probability compared with existing schemes. {We leave some further analysis, such as the average power consumption of D2D UEs and the impact of backoff window on the average delay of D2D UEs, for future work.}

\bibliographystyle{IEEEtran}
\bibliography{IEEEabrv,D2D_discovery}

\end{document}